\documentclass[aps,prl,twocolumn,superscriptaddress]{revtex4}

\usepackage[dvips]{graphicx}
\usepackage{amsmath}
\usepackage{amsfonts}
\usepackage{amssymb}
\usepackage[latin1]{inputenc}
\usepackage{xcolor}

\definecolor{darkblue}{rgb}{0, 0, 0.8}
\usepackage[colorlinks=true, breaklinks=true, linkcolor=darkblue, citecolor=darkblue, urlcolor=darkblue]{hyperref}
\newcommand{\doilink}[2]{\href{http://dx.doi.org/#1}{#2}}

\begin{document}

\title{Coherent scattering of near-resonant light by a Dense Microscopic Cold Atomic cloud}

\author{S. Jennewein}
\affiliation{Laboratoire Charles Fabry, Institut d'Optique, CNRS, Univ Paris Sud,
2 Avenue Augustin Fresnel,
91127 Palaiseau cedex, France}
\author{M. Besbes}
\affiliation{Laboratoire Charles Fabry, Institut d'Optique, CNRS, Univ Paris Sud,
2 Avenue Augustin Fresnel,
91127 Palaiseau cedex, France}
\author{N.J. Schilder}
\affiliation{Laboratoire Charles Fabry, Institut d'Optique, CNRS, Univ Paris Sud,
2 Avenue Augustin Fresnel,
91127 Palaiseau cedex, France}
\author{S.D. Jenkins}
\affiliation{Mathematical Sciences, University of
  Southampton, Southampton SO17 1BJ, United Kingdom}
\author{C. Sauvan}
\affiliation{Laboratoire Charles Fabry, Institut d'Optique, CNRS, Univ Paris Sud,
2 Avenue Augustin Fresnel,
91127 Palaiseau cedex, France}
\author{J. Ruostekoski}
\affiliation{Mathematical Sciences, University of
  Southampton, Southampton SO17 1BJ, United Kingdom}
\author{J.-J. Greffet}
\affiliation{Laboratoire Charles Fabry, Institut d'Optique, CNRS, Univ Paris Sud,
2 Avenue Augustin Fresnel,
91127 Palaiseau cedex, France}
\author{Y.R.P. Sortais}
\affiliation{Laboratoire Charles Fabry, Institut d'Optique, CNRS, Univ Paris Sud,
2 Avenue Augustin Fresnel,
91127 Palaiseau cedex, France}
\author{A. Browaeys}
\affiliation{Laboratoire Charles Fabry, Institut d'Optique, CNRS, Univ Paris Sud,
2 Avenue Augustin Fresnel,
91127 Palaiseau cedex, France}

\date{\today}

\begin{abstract}
We measure the coherent scattering of light by a cloud of laser-cooled atoms with a
size comparable to the wavelength of light. By interfering a laser beam tuned near an atomic
resonance with the field scattered by the atoms we observe a resonance with a
red-shift,
a broadening, and a saturation of the extinction for increasing
atom numbers. We attribute these features to enhanced light-induced dipole-dipole interactions in a cold, dense
atomic ensemble that result in a failure of standard predictions such as the ``cooperative Lamb shift''.
The description of the atomic cloud by a mean-field model based on the Lorentz-Lorenz
formula that ignores scattering events where light is scattered recurrently by the same atom and
by a microscopic discrete dipole model that incorporates these effects lead to progressively closer
agreement with the observations, despite remaining differences.
\end{abstract}
\pacs{42.50.Ct,42.50.Nn,42.25.Fx,32.80.Qk,03.65.Nk}

\maketitle

The understanding of  light propagation in dense media relies traditionally
on a continuous description
of the sample characterized by macroscopic
quantities such as susceptibility or refractive index~\cite{Jackson,BornandWolf}.
Their derivation from a microscopic
theory is in general challenging owing to the interactions between the light-induced dipoles
that can be large when the light is tuned near an atomic resonance.
In dilute media, their role can be analyzed using the perturbative
approach of Friedberg, Hartmann and Manassah (FHM)~\cite{Friedberg1973},
which predicts in particular a  ``cooperative Lamb-shift'' measured recently in
inhomogeneously broadened media~\cite{Keaveney2012a,Rohlsberger2010}
and cold dilute atomic gases~\cite{Roof2016}.
For an atom slab, the FHM approach was shown to correspond to
the low-density limit of
the local-field model introduced by Lorentz~\cite{Javanainen2016}, which
replaces the action of all the atoms of the medium on a particular one
by an average effective field~\cite{Jackson,BornandWolf},
thus ignoring correlations between
the light-induced dipoles.
This mean-field approach leads to the Lorentz-Lorenz formula, which allows
calculating the index of refraction of many dense media
with an excellent accuracy~\cite{BornandWolf,RMPSchnatterly1992}.
However it was pointed out~\cite{Javanainen2014a,Javanainen2016} that in the absence
of inhomogeneous broadening, such as in cold atomic ensembles,
the mean-field response may not be valid due to recurrent scattering where
the field radiated by one atom can be scattered back by another
atom~\cite{Morice1995,Ruostekoski1997}.
Recurrent scattering should become important when
the incident light (wavelength $\lambda=2\pi/k$)
is tuned near an atomic
resonance, and the atomic density approaches $k^3$.
This calls for an experiment operating in this regime, where a  comparison between the
standard mean-field theories of light scattering and a microscopic approach
including recurrent scattering can be performed.

Here, we perform this comparison. To do so we need to access
a quantity relevant to both the macroscopic
and the microscopic approaches.
The {\it coherent}  electric field $\langle {\bf E}_{\rm sc}\rangle$ scattered
by the cloud fulfills this condition: it is obtained by averaging the
scattered field ${\bf E}_{\rm sc}$
over many realizations of the
spatial random distribution of atoms, and its evolution
is governed by the macroscopic
Maxwell's equations in the cloud considered as an homogeneous
medium described by a susceptibility.
In the case of cold atomic gases, the near-resonance coherent optical response
has been explored experimentally using mostly  dilute, optically thick ensembles~\cite{Bienaime2010,Bender2010,Chabe2014,Chalony2011,Kwong2014,
Roof2014,Guerin2016,Bromley2016}.
Recently, we studied the light scattered by a microscopic {\it dense} cloud of cold
atoms at $90^\circ$ of a near-resonant excitation laser~\cite{Pellegrino2014}.
In that situation, we were sensitive to the {\it incoherent} component
$\langle |{\bf E}_{\rm sc}-\langle {\bf E}_{\rm sc}\rangle|^2\rangle$ of the scattered light.
We could therefore not compare our results with mean-field
predictions for continuous media, which are only relevant
for the coherent part.

In this work, we study the {\it coherent} scattering by our microscopic cloud.
The cloud contains up to a few hundreds laser-cooled rubidium 87 atoms and has a
size smaller than the wavelength of the optical D2 transition.
We illuminate the sample with a tightly focused laser with a waist larger than the cloud size.
We access the coherent scattering by measuring the extinction resulting from the interference
of the laser field with the field scattered by the cloud.
We observe a saturation of the extinction,
a broadening of the line, and a small red-shift when we vary the number of atoms from 10 to 180.
We show that the measured shift and width do not agree with the FHM perturbative
approach. The description of the atomic cloud by a mean-field model based on the Lorentz-Lorenz
formula also disagrees with our data. Finally a microscopic discrete dipole model
that incorporates recurrent scattering leads to a qualitatively closer
agreement with our measurements, despite remaining differences. 

\begin{figure}
\includegraphics[width=\columnwidth]{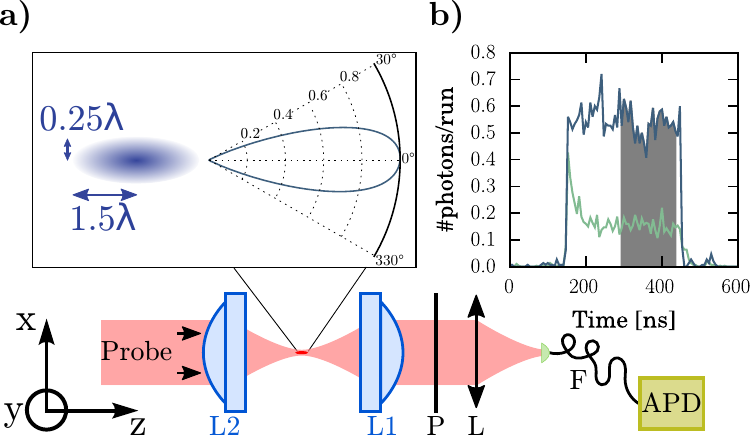}
\caption{(a) Experimental setup. A microscopic cloud of $^{87}$Rb atoms is illuminated by a linearly-polarized probe laser focused down to a waist $w=1.2\,\mu$m. (P): polarizer. (L)  Lens allowing the mode-matching between the laser probe beam and the single-mode fiber (F) in the absence of atoms. (APD): avalanche photo-diode. Inset: cloud rms widths (left) and intensity radiation pattern (right), calculated using a microscopic approach (see text). The coherent part, $|{\cal E}_{\rm coh}(\omega)|^2$, dominates the incoherent part, which is more isotropic and is orders of magnitude smaller. (b) Example of temporal signals recorded on the APD with $N=180$ atoms (green line) and without atoms (blue line). The laser is nearly resonant with the atoms, with a frequency detuning $\Delta = 0.3\Gamma$. Each run consists of $1000$ illuminations with a duration of $300$\,ns each. Temporal bins: $6$\,ns. Grey area: time interval used for the steady-state analysis. }
\label{fig1_setup}
\end{figure}

To study the coherent scattering by our cloud, we detect the interference in the far field between the laser field ${\bf E}_{\rm L}$ and the scattered field ${\bf E}_{\rm sc}$. To do so, we use two identical aspherical lenses L1 and L2 with a high numerical aperture (NA=$0.5$) mounted in a confocal configuration in a vacuum chamber (see Fig.\,\ref{fig1_setup}a) \cite{Sortais2007}. L1 focuses far-off-detuned laser light onto a waist of $1.2\pm0.1\,\mu$m ($1/e^2$ radius). This creates a dipole trap (depth: $1$\,mK) in which we load $N$ atoms with a temperature of $120\pm15\,\mu$K~\footnote{The Doppler
broadening  ($k\Delta v=0.06\Gamma$) is negligible with respect to the linewidth $\Gamma$.}. We control the number of atoms $N$ within 10\%, and vary $N$ between 10 and 180~\footnote{The exact number of atoms inside the trap varies shot to shot as it is governed by a sub-Poisson distribution with a mean $N$ and variance $3N/4$~\cite{Sortais2012}.}. The atomic cloud is cigar-shaped, with calculated transverse and longitudinal root-mean-square (rms) widths $(a_\perp, a_z)=(0.2, 1.2)\,\mu$m. The peak densities range from $n=10^{13}$ to $2\times 10^{14}$ at/${\rm cm}^3$. The uncertainties in the temperature, atom number and waist size lead to a
systematic uncertainty on the peak density of a factor $2$.
The probe beam is focused down to a waist of $w=1.20\pm0.05\,\mu$m
also by L2 at the position of the cloud.
It is linearly polarized and nearly resonant with the closed D2 transition of rubidium between the $(5S_{1/2},F=2)$ and $(5P_{3/2},F=3)$ levels at $\lambda =2\pi c/\omega_0= 780.2$\,nm (linewidth $\Gamma = 2\pi\times 6$\,MHz)~\footnote{The magnetic field is compensated to values lower than $\sim 80$\,mG.}. We operate the probe in the low  intensity limit where the atoms respond linearly to the field: $I/I_{\rm sat}\approx 0.04$ ($I_{\rm sat}=1.6$\,mW/cm$^2$).
We collect the probe light transmitted through the cloud using L1 and couple it
into a single-mode fiber connected to an avalanche photodiode (APD). The temporal signals are acquired by accumulating single photons using a counting card with a resolution of $150$\,ps. A polarization beam-splitter is placed before the single-mode fiber and aligned at $45^\circ$ of the probe laser polarization so as to split the collected light between the fibered APD and a CCD camera (not shown in Fig.\,\ref{fig1_setup}a).

Our configuration is sensitive to the mode-matching ${\cal E}(\omega)= \int \left\{{\bf E}({\bf r},\omega)\cdot {\bf g}^*({\bf r})\right\}dS$ between the total field ${\bf E}={\bf E}_{\rm L}+{\bf E}_{\rm sc}$ and the mode ${\bf g}$ of the single-mode-fibered detector ($dS$ is a differential area element perpendicular to the optical axis)~\cite{Abdullah2009}. In the absence of atoms the fiber mode is matched to the incoming light, i.e. ${\bf g}\propto {\bf E}_{\rm L}$.  In our experiment, we measure $\langle |{\cal E}(\omega)|^2\rangle$, where $\langle \cdot \rangle$ means an average over many realizations of the cloud. After averaging, the signal is the sum of two parts~\cite{SM}: (i) $|{\cal E}_{\rm coh}(\omega)|^2$ due to ${\bf E}_{\rm L}+\langle {\bf E}_{\rm sc}\rangle$, and (ii)
$\langle |{\cal E}_{\rm incoh}(\omega)|^2\rangle$ due to the fluctuating
field ${\bf E}_{\rm sc}-\langle {\bf E}_{\rm sc}\rangle$.
In the direction of propagation of the laser $|{\cal E}_{\rm coh}(\omega)|^2 \gg \langle |{\cal E}_{\rm incoh}(\omega)|^2\rangle$ (see below and in Fig.~\ref{fig1_setup}a) and we are therefore mainly sensitive to the coherent optical response, which we characterize by a transfer function ${\cal S}(\omega)=\langle{\cal E}(\omega)\rangle/{\cal E}_{\rm L}(\omega)$ obtained by comparing the detected fields with and without atoms.

\begin{figure}
\includegraphics[width=\columnwidth]{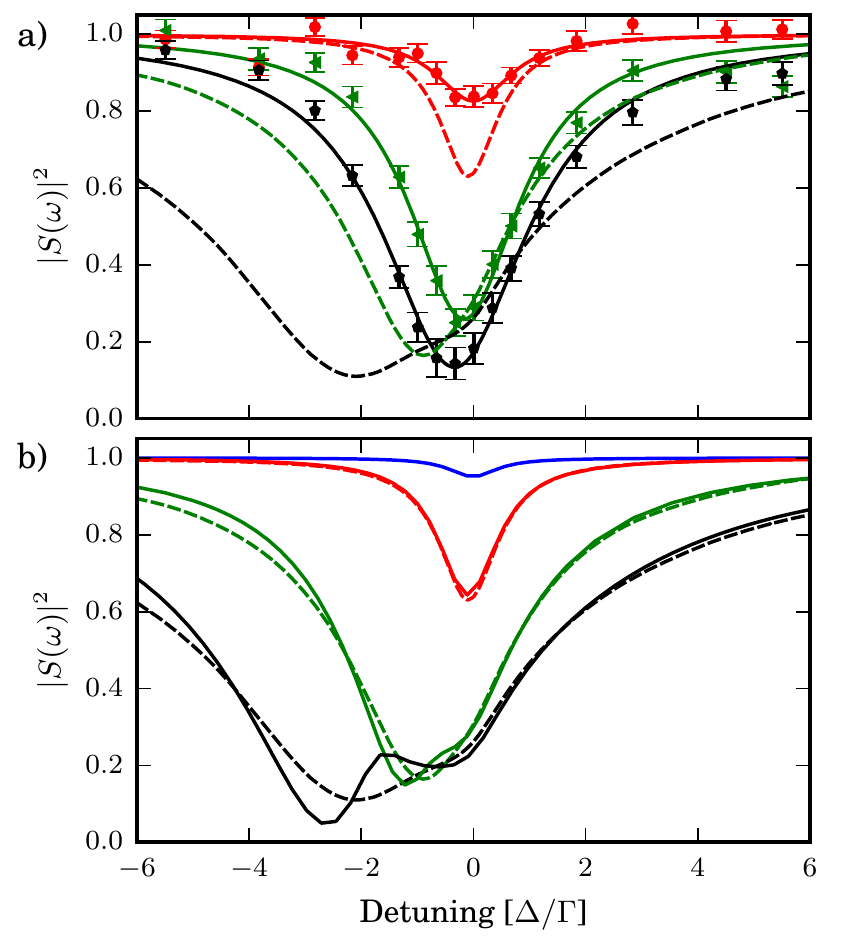}
\caption{(a) Measured transfer function of the cloud in steady-state versus probe detuning $\Delta$ for $N=(10,83,180)$ atoms (top to bottom); error bars: statistical (one standard deviation), shot noise limited. Solid lines: Lorentzian fit by $|{\cal S}(\omega)|^2$.
Dotted lines: results of the coupled dipole equations including the 12 levels of the $F_{\rm g}=2-F_{\rm e}=3$ transition (see text).
(b)  Comparison between the predictions of the Lorentz model
(solid line) and the
microscopic, 12-level atom model (dotted line) for $N=(1,10,83,180)$ (top to bottom).}
\label{fig2_Transmission}
\end{figure}

To measure ${\cal S}(\omega)$ in steady-state, we proceed in the following way: after preparing the atoms in the $(5S_{1/2},F=2)$ level, we switch off the dipole trap light during $500$\,ns and send a $300$\,ns probe pulse with a temporal top hat profile (rise time of $2$\,ns). We then recapture the cloud in the trap for $500$\,ns and repeat this release-probe-recapture 1000 times using the same atomic cloud~\footnote{Our results vary by less than $5\%$ when the same cloud is illuminated with a single shot probe. This observation rules out a possible cumulative modification of the cloud volume during the pulsed excitation.}.
This procedure is typically repeated with $200$ different atomic clouds.
A typical signal is shown in Fig.\,\ref{fig1_setup}b. It reaches a steady-state after a transient time of $\sim 26$\,ns, close to the lifetime $1/\Gamma$ of the excited state, during which the atomic medium gets polarized. We average the signal over a time interval of $120$\,ns (grey area) and normalize it with respect to the case without atoms to obtain the transmission in steady state for a given probe frequency. We checked that the scattered light has the same polarization as the probe light by rotating the polarizer $P$ and observing a signal with a contrast of $95\%$, the same as in the absence of atoms. This feature is characteristic of the coherent scattered field, and therefore confirms experimentally that $|{\cal E}_{\rm coh}(\omega)|^2 \gg \langle |{\cal E}_{\rm incoh}(\omega)|^2\rangle$.
Finally, the sequence is repeated for various probe detunings $\Delta=\omega-\omega_0$ and atom numbers $N$. We obtain the spectra shown in Fig.\,\ref{fig2_Transmission}a.

\begin{figure}
\includegraphics[width=\columnwidth]{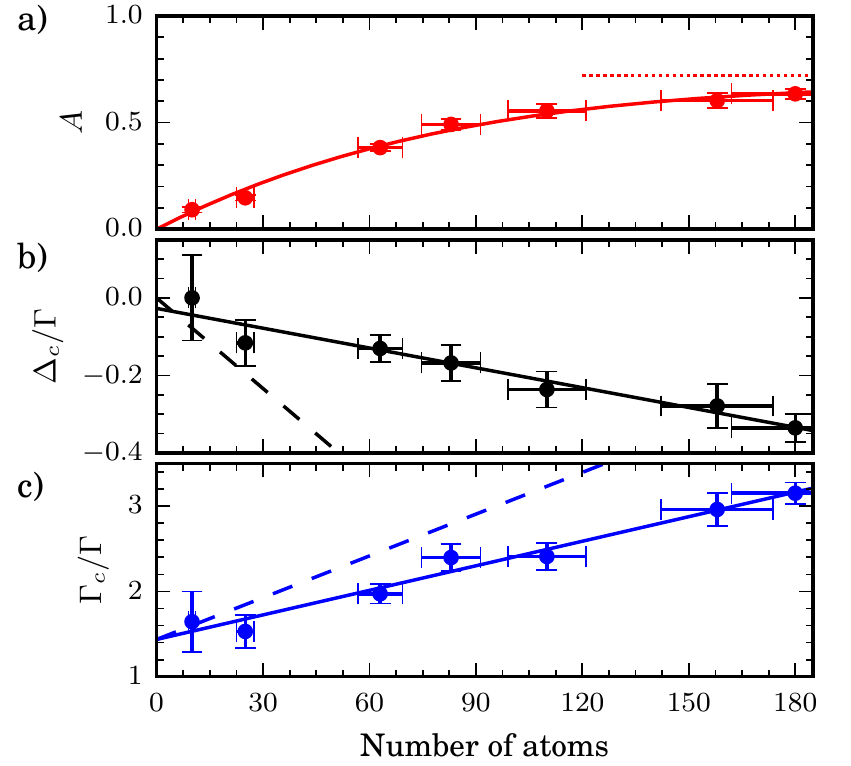}
\caption{Fit results of the data of Fig.\,\ref{fig2_Transmission} with the function $|{\cal S}(\omega)|^2$. Error bars are from the fit.
(a) Amplitude $A$. Solid green line: phenomenological fit to guide the eye, yielding a saturation (dotted line) at $0.7$.
(b) Shift of the center-frequency $\Delta_c=\omega_c-\omega_0$. Solid line: linear fit.
(c) Full width at half maximum $\Gamma_c$. Solid line: linear fit.
Dashed lines in (b) and (c): predictions by Friedberg, Hartmann and Manassah detailed in the text.
The prediction for the width
has been offset to match the data for $N=0$.}
\label{Fig3_summaryabsorption}
\end{figure}

The derivation of a functional form for ${\cal S}(\omega)$ is very hard in our dense cloud regime.
However in the case of a cloud with a size smaller than $1/k$ so that it
behaves as a small dielectric sphere
with a polarizability $\alpha_{\rm c}(\omega)$ we get, following~\cite{Abdullah2009}, ${\cal S}(\omega)\approx1 + i k\alpha_{\rm c}(\omega)/(\pi w^2)$, which we cast in the form:
\begin{equation}\label{Eq:S(omega)}
{\cal S}(\omega) = 1-{A\over 1-2i{\omega-\omega_c \over \Gamma_c}} \ ,
\end{equation}
assuming that the polarizability is resonant around a frequency $\omega_{\rm c}$
with a width $\Gamma_{\rm c}$.
We fit the spectra shown in Fig.~\ref{fig2_Transmission}a with the Lorentzian function $|{\cal S}(\omega)|^2$, using Eq.~(\ref{Eq:S(omega)}) and leaving $A$, $\Delta_c=\omega_c-\omega_0$ and $\Gamma_c$ as free parameters. The fit agrees well with the data, confirming that the functional form of Eq.\,(\ref{Eq:S(omega)}) is appropriate even for our elongated sample. Figure\,\ref{Fig3_summaryabsorption} shows the results of the fits. For increasing atom numbers, we observe a saturation of the amplitude $A$, and therefore of the extinction, an increasing small red-shift and a broadening of the line. These behaviors can be understood qualitatively as a consequence of the dipole-dipole interactions between atoms, on the order of $\hbar \Gamma n/k^3$ (see below). 

We now  compare our results to various models of the optical response of the cloud.
In Refs.~\cite{Friedberg1973,Manassah2012}, Friedberg, Hartmann and Manassah used perturbation theory
to derive the expressions for a collective decay rate and a collective shift for various geometries of an atomic ensemble of two-level atoms.  There, the collective shift and rate
are the real and imaginary parts of the average dipole-dipole interaction~\cite{Scully2009}. This theory predicts  the ``cooperative Lamb-shift'' measured  in a hot atomic vapor~\cite{Keaveney2012a} and in a dilute, optically thick cold atomic sample~\cite{Roof2016}.
For the case of an ellipsoidal cloud with Gaussian density distribution,
the predictions (see formulae (5.2) and (5.3) of Ref.~\cite{Manassah2012})
are plotted in Fig.\,\ref{Fig3_summaryabsorption}b,c for our experimental parameters.
Here, we included the rubidium internal structure
by multiplying the prediction of Ref.~\cite{Manassah2012}
by the ratio of multiplicities 7/15 of the $F_{\rm g}=2-F_{\rm e}=3$
transition~\cite{Ruostekoski2000,Mueller2001},
assuming equal populations in all hyperfine Zeeman ground states and a negligible
magnetic field (as is the case in the experiment)~\cite{SM}.
The predictions
differ significantly from the measured values,
indicating that this perturbative approach does not apply for our dense, cold atomic systems.

To go beyond the FHM perturbative treatment,
we now calculate the optical response as predicted by the Lorentz local field theory for our {\it dense}
cigar-shaped cloud.
For this purpose, we replace the cloud by a Gaussian continuous density distribution $n({\bf r})$ (with rms widths $a_\perp$ and $a_z$) and calculate the local susceptibility using the Lorentz-Lorenz formula $\chi({\bf r},\omega) = n({\bf r})\alpha(\omega)/(1-n({\bf r})\alpha(\omega)/3)$~\cite{Jackson,BornandWolf}.
Here $\alpha(\omega)= i(7/15)(6\pi /k^3)/[1-i (2\Delta/\Gamma)]$ is the polarizability of a
single atom, which includes the internal atomic structure of rubidium as described above, see~\cite{SM}.
We then define a local permittivity  $\epsilon({\bf r})=1+\chi({\bf r})$ and use a
finite element program to calculate the electric field scattered in the far field by the cloud  illuminated
by the Gaussian laser beam. We finally compute the transfer function ${\cal S}(\omega)$
taking for the Gaussian field the usual paraxial expression~\cite{SM}.
The results are shown in Fig.\,\ref{fig2_Transmission}b.
The mean-field response of the cloud predicted by the Lorentz-Lorenz formula deviates from the data as the number of atoms increases,
featuring in particular a double structure for the largest atom numbers~\footnote{The double structure
appearing in the Lorentz model indicates that
the resonance of the coherent response of the cloud
is not only related to the resonance of the susceptibility,
but also to a  shape resonance.}, as well as a large asymmetry
(also observed in the spectrum of transmitted light of an atomic slab
described by the Lorentz-Lorenz formula~\cite{Javanainen2016}).

We finally calculate the coherent response of the cloud using a microscopic model
where the atoms are considered as point-like dipoles ${\bf d}_j$ randomly positioned
according to the Gaussian spatial distribution, each
being driven by the laser field and the fields scattered by all the
other ones~\cite{Ruostekoski1997,Chomaz2012}.
This approach leads to a set of coupled dipole equations.
As in Ref.~\cite{Pellegrino2014},
we include the internal structure of the atoms by randomly assigning them a given Zeeman state  $m_j$ of the $(5S_{1/2},F=2)$ manifold and we write ${\bf d}_j=\mathcal{D}\sum_{\sigma}{\bf\hat{e}}_\sigma C^{(\sigma)}_{m_j}{\cal P}_{j\sigma}$ ($\sigma=\pm1,0$ defines the polarization).
The amplitude of the atomic dipole $j$ associated to the optical
transition $|g,m_j\rangle\rightarrow |e,m_j+\sigma\rangle$ is proportional to the
reduced dipole matrix element $\mathcal{D}$, the atomic coherence
${\cal P}_{j\sigma}$, and the corresponding Clebsch-Gordan coefficient $C^{(\sigma)}_{m_j}$.
We solve the steady state set of coupled equations for the coherences
\begin{equation}\label{Eq:coupled_dipoles}
\left(\Delta+i\Gamma/2\right) {\cal P}_{j\alpha}
=\Omega_{j\alpha} + \sum_{l\neq j}\sum_{\beta} C^{(\beta)}_{m_l}C^{(\alpha)}_{m_j}{V_{j\alpha}^{l\beta}({\bf r})}\, {\cal P}_{l\beta}
\ ,\nonumber
\end{equation}
where $V_{j\alpha}^{l\beta}=-V_{\rm dd}\left[ p_{\alpha\beta}(i k r -1) + q_{\alpha\beta}(kr)^2\right]e^{i kr}$
with $V_{\rm dd}=3\hbar\Gamma/4(k r)^3 $ is the dipole-dipole interaction, $p_{\alpha\beta}$ and $q_{\alpha\beta}$ are angular functions~\cite{Pellegrino2014}, and $\Omega_{j\alpha}$ the Rabi frequency.
We calculate the field scattered by the cloud (yielding the radiation pattern shown in Fig.\,\ref{fig1_setup}a). We then compute the interference of this field with the laser field, at the position of the lens L1, and the transfer function for this particular configuration of the atomic ensemble
and average over many spatial configurations.

The results of the microscopic model are plotted in Fig.\,\ref{fig2_Transmission}b for various detunings and atom numbers, together with the prediction of the Lorentz local field model. We observe that both models are in  agreement for low values of $N$, and predict approximately Lorentzian
lineshapes. For large atom numbers, however, they
differ quantitatively, pointing towards the role of
recurrent scattering, included in the microscopic model, to all orders, but not in the Lorentz 
model~\cite{Morice1995,Ruostekoski1997,Lagendijk1996,Javanainen2014a,Sokolov2009}.
To the lowest order in density, for a cloud (density $n$) of identical atoms
with polarizability $\alpha$,
the contribution of recurrent scattering to the susceptibility is proportional to
the number of atom pairs $(n\alpha)^2$ inside the scattering volume $\alpha$. It
becomes important when
$n\alpha \sim 1$.
The onset of light-induced correlations and the effect of recurrent scattering 
as a function of the detuning and atom density was analyzed in more detail in~\cite{Saunders1973,Morice1995,Ruostekoski1997,Ruostekoski1999,Javanainen2016}.
In the presence of recurrent scattering and when $n/k^3\ll1$,
the susceptibility takes the form:
\begin{equation}\label{Eq:LLmodifie}
\chi(\omega)={n\alpha(\omega) \over 1-n\alpha(\omega)({1\over 3}+\beta(\omega))}\ ,
\end{equation}
where $\beta(\omega)$ is the contribution from recurrent scattering.
Using formula (22) of~\cite{Morice1995},
we get $\beta(\omega)\propto B \alpha(\omega) k^3$,
with $B$ a volume integral challenging to calculate for our geometry. The 
lowest order contribution to $\beta$ is independent of the density.
If $B\sim 1$, the local field correction $n\alpha/3$ is thus on the
same order as the recurrent scattering contribution 
close to resonance ($\alpha=6\pi i/k^3$),  while
away from resonance ($\alpha k^3\ll 1$) the influence of recurrent scattering is negligible.
Remarkably, when inhomogeneous broadening is introduced
(such as the Doppler effect in hot vapor cells~\cite{Keaveney2012a}),
the resonant frequencies of the dipoles $j$ are spread over $\Delta\omega_{\rm D}$
and $\beta$ is replaced by the average over $j$,
$\langle \alpha\rangle_j k^3$, and is therefore reduced by
a factor $\Gamma/\Delta\omega_{\rm D}$~\cite{Javanainen2014a}.
This explains why for any medium where inhomogeneous broadening  is dominant
the Lorentz-Lorenz model is successful, as $\Gamma/\Delta\omega_{\rm D}\ll 1$
and thus $\beta\ll 1$. On the contrary, in the absence of inhomogeneous broadening,
the Lorentz-Lorenz formula is usually not valid at resonance.
In~\cite{Schilder2015}, we use the microscopic approach  
to calculate the effective dielectric constant of our cloud, but in the regime $n/k^3\ge 1$, 
and found that it does not follow the Lorentz-Lorenz formula, as expected.

Finally, we compare our measurements to the microscopic model
(see Fig.\,\ref{fig2_Transmission}a). We observe that the data are closer to this model than
to the Lorentz  model, as they do not show the double structure predicted by the Lorentz model
for the largest atom numbers. This  indicates that the Lorentz model is not valid in our configuration.
However the measurements exhibit systematically
less pronounced features for the shift, width and amplitude than predicted by the microscopic model.
On the experimental side, we have ruled out possible biases, 
such as the probe beam alignment~\cite{SM}
and the possible cumulative heating of the cloud due to the pulsed
illumination that could result in a modification of the cloud volume.
On the theoretical side, the models ignore quantum fluctuations between hyperfine ground states and
assume the low light intensity limit, which may in practice be
difficult to fully realize in the experiments due to secondary radiation by closely-spaced atoms.

As a conclusion, we have measured the coherent scattering by a dense,  
cold atomic cloud. We have observed a
failure of standard models, such as the FHM model or 
the mean-field Lorentz model. The remaining difference
with the microscopic model shows that a quantitative understanding of the light-induced interactions 
even in a relatively simple situation is still a challenge.

\begin{acknowledgments}
We thank C.S.~Adams for discussions.
We acknowledge support from the E.U. (ERC Starting Grant ARENA
and the HAIRS project), from the Triangle de la Physique (COLISCINA project),
the labex PALM (ECONOMIC project)
and the Region Ile-de-France (LISCOLEM project), the EPSRC, and the Leverhulme Trust.
N.J. Schilder is supported by Triangle de la Physique.
JJG is a senior member of the Institut Universitaire de France.
\end{acknowledgments}

\end{document}


\title{Supplemental Material: Coherent scattering of near-resonant light by a Dense Microscopic Cold Atomic cloud}

\author{S. Jennewein}
\affiliation{Laboratoire Charles Fabry, Institut d'Optique, CNRS, Univ Paris Sud,
2 Avenue Augustin Fresnel,
91127 Palaiseau cedex, France}
\author{M. Besbes}
\affiliation{Laboratoire Charles Fabry, Institut d'Optique, CNRS, Univ Paris Sud,
2 Avenue Augustin Fresnel,
91127 Palaiseau cedex, France}
\author{N.J. Schilder}
\affiliation{Laboratoire Charles Fabry, Institut d'Optique, CNRS, Univ Paris Sud,
2 Avenue Augustin Fresnel,
91127 Palaiseau cedex, France}
\author{S.D. Jenkins}
\affiliation{Mathematical Sciences, University of
  Southampton, Southampton SO17 1BJ, United Kingdom}
\author{C. Sauvan}
\affiliation{Laboratoire Charles Fabry, Institut d'Optique, CNRS, Univ Paris Sud,
2 Avenue Augustin Fresnel,
91127 Palaiseau cedex, France}
\author{J. Ruostekoski}
\affiliation{Mathematical Sciences, University of
  Southampton, Southampton SO17 1BJ, United Kingdom}
\author{J.-J. Greffet}
\affiliation{Laboratoire Charles Fabry, Institut d'Optique, CNRS, Univ Paris Sud,
2 Avenue Augustin Fresnel,
91127 Palaiseau cedex, France}
\author{Y.R.P. Sortais}
\affiliation{Laboratoire Charles Fabry, Institut d'Optique, CNRS, Univ Paris Sud,
2 Avenue Augustin Fresnel,
91127 Palaiseau cedex, France}
\author{A. Browaeys}
\affiliation{Laboratoire Charles Fabry, Institut d'Optique, CNRS, Univ Paris Sud,
2 Avenue Augustin Fresnel,
91127 Palaiseau cedex, France}

\date{\today}

\maketitle

This Supplemental Material presents more details about the definition of the coherent
transfer function ${\cal S}(\omega)$
and the expressions of the laser field used in the modeling.
It also derives the mean-field expression of the susceptibility for a multi-level alkali atom
that we use in the Lorentz-Lorenz formula to calculate the field scattered by the cloud.
Finally, we discuss the influence on the line shape
of a misalignment of the laser probe
with respect to the atomic cloud.

\section{Coherent transfer function}

The experimental configuration used in the experiment gives access to
the overlap between the total field in the forward direction ${\bf E}={\bf E}_{\rm L}+{\bf E}_{\rm sc}$
and the mode ${\bf g}$ of the single-mode-fibered detector
\begin{equation}\label{Eq:ModeMatching}
{\cal E}(\omega)= \int {\bf E}({\bf r},\omega)\cdot {\bf g}^*({\bf r})dS
\end{equation}
with $dS$ a differential area element perpendicular to the optical axis.
As the fiber mode is matched to the incoming light, ${\bf g}\propto {\bf E}_{\rm L}$, and
the total transfer
function is
\begin{equation}\label{Eq:TransferS}
{\cal S}_{\rm tot}(\omega) = {{\cal E}(\omega)\over {\cal E}_{\rm L}(\omega)}=
{\int {\bf E}({\bf r},\omega)\cdot {\bf E}_{\rm L}^*({\bf r})dS \over  \int  |{\bf E}_{\rm L}({\bf r})|^2 dS}\ .
\end{equation}
We then decompose the scattered field into the coherent and incoherent (fluctuating) components:
${\bf E}_{\rm sc} = \langle {\bf E}_{\rm sc}\rangle+\delta {\bf E}_{\rm sc}$,
where $\langle \cdot \rangle$ indicates an average over many spatial configurations of the cloud.
In the experiment, we measure the configuration-averaged quantity
\begin{eqnarray}
\langle|{\cal E}(\omega)|^2\rangle&\propto&
\left| \int ({\bf E}_{\rm L}+\langle {\bf E}_{\rm sc}\rangle)\cdot {\bf E}_{\rm L}^*dS \right|^2  \\
&+&\langle\left|\int \delta{\bf E}_{\rm sc}\cdot {\bf E}_{\rm L}^*dS  \right|^2\rangle \ ,\nonumber
\end{eqnarray}
(taking into account $\langle\delta{\bf E}_{\rm sc}\rangle=0$), which we cast in the form
$|{\cal E}_{\rm coh}(\omega)|^2+\langle |{\cal E}_{\rm incoh}(\omega)|^2\rangle$. As explained in the main text,
$|{\cal E}_{\rm coh}(\omega)|^2 \gg \langle |{\cal E}_{\rm incoh}(\omega)|^2\rangle$
in the direction of propagation of the laser, and therefore, we measure essentially
the coherent part to the total transfer function. We thus define the coherent
optical transfer function:
\begin{equation}\label{Eq:TransferS}
{\cal S}(\omega)={\langle{\cal E}(\omega)\rangle\over {\cal E}_{\rm L}(\omega)}=
1+{\int \langle {\bf E}_{\rm sc}\rangle\cdot {\bf E}_{\rm L}^*dS \over  \int  |{\bf E}_{\rm L}|^2 dS}\ .
\end{equation}

In general, the transfer function defined above does not correspond
to the transmission of the cloud defined as
\begin{equation}\label{Eq:TransmissionT}
T(\omega)={\langle\int |{\bf E}({\bf r},\omega)|^2 dS \rangle\over  \int  |{\bf E}_{\rm L}({\bf r})|^2 dS}\  .
\end{equation}
If the solid angle of the collecting lens L1 is very small, $T(\omega)$ and $|{\cal S}_{\rm tot}(\omega)|^2$ coincide.
Otherwise,
the relation between the transfer function and the transmission
can be found by decomposing, once again, the total field as
${\bf E}={\bf E}_{\rm L}+\langle {\bf E}_{\rm sc}\rangle+\delta {\bf E}_{\rm sc}$. We then get
\begin{eqnarray}
T(\omega)&=&|{\cal S}(\omega)|^2+{\int \langle|\delta {\bf E}_{\rm sc}|^2\rangle dS \over  \int  |{\bf E}_{\rm L}({\bf r})|^2 dS}\\
&+&{\int |\langle {\bf E}_{\rm sc}\rangle|^2 dS \over  \int  |{\bf E}_{\rm L}({\bf r})|^2 dS}-\left|{\int\langle {\bf E}_{\rm sc}\rangle\cdot {\bf E}_{\rm L}^* dS \over  \int  |{\bf E}_{\rm L}({\bf r})|^2 dS}\right|^2\ . \nonumber
\end{eqnarray}
Using the Cauchy-Schwartz inequality
$$\left|\int\langle {\bf E}_{\rm sc}\rangle\cdot {\bf E}_{\rm L}^* dS \right|^2
\le \int |\langle {\bf E}_{\rm sc}\rangle|^2 dS \int  |{\bf E}_{\rm L}({\bf r})|^2 dS$$
yields $|{\cal S}(\omega)|^2\leq T(\omega) $.

\section{Description of the probe laser field}

In the theoretical models  (Lorentz local-field and microscopic),
we use the following paraxial approximation for the
amplitude of the $x$ component of the electric  probe laser field (with a waist $w$, see Fig.~1a in main text):
\begin{equation}\label{Eq:Gauss_beam}
E_{{\rm L}, x}(x,y,z)={E_0\over 1+i{z\over z_{\rm R}}} \exp\left[i k{x^2+y^2 \over 2 q(z)}\right]\, \exp[ikz]\ ,
\end{equation}
with $\frac{1}{q(z)}=\frac{1}{R(z)}+\frac{2\,\mathrm{i}}{kw^2(z)}$, $z_{\rm R} = kw^2/2$ the Rayleigh length, $w(z)=w\sqrt{1+z^2/z_{\rm R}^2}$
and $R(z)=z+z_{\rm R}^2/z$.

As is well-known, this paraxial expression is not an exact solution of Maxwell's equation. We have therefore checked, for the calculation of the transfer function based on the Lorentz-Lorenz formula, that the
field scattered by the cloud using the paraxial approximation is numerically very close to the one
obtained by using the plane wave decomposition of the incident beam (which corresponds to an exact solution of Maxwell's equations).

\section{Derivation of the mean-field susceptibility for a multi-level atom}

In this section we derive the mean-field electric susceptibility for a multi-level atom, when the recurrent scattering
contributions are ignored. This is then used in the main section of the paper to calculate the mean-field response
and the ``cooperative Lamb shift'' for a $^{87}$Rb $F=2$ ground-state manifold that differ from the ones obtained for
an isotropic $F=0\rightarrow F'=1$ transition.

We use the general formalism of Ref.~\cite{Ruostekoski1997} and derive the multi-level electric susceptibility as
in Ref.~\cite{Ruostekoski2000}. We consider an atom with ground states $ |g,\nu\>$ and excited states $|e,\eta\>$. Here
$\nu$ and $\eta$ represent the Zeeman sub-levels of the ground and excited states, respectively,
separated
by a transition at a frequency $\omega_0$.

The positive frequency component of the electric field amplitude is the sum of the incident coherent field $ \Dv^+_F({\bf r}) $
and the scattered field from the atomic polarization $\Pvhat^+(\rv)$,
\beq
\eo\Evhat^+({\bf r}) = \Dv^+_F({\bf r}) +
\int d^3r'\,
{\sf G}({\bf r}-{\bf r'})\,\Pvhat^+({\bf r}')\,.
\label{eq:MonoD}
\eeq
The monochromatic dipole radiation kernel
${\sf G}(\rv)$~\cite{Jackson} gives the radiated field at $\rv$ from
a dipole with the amplitude $\hat{\bf d}$ residing at the origin:
\begin{align}
{\sf G}&({\bf r})\,\hat{\bf d}=
{k^3\over4\pi}
\left\{ (\hat{\bf n}\!\times\!\hat{\bf d}
)\!\times\!\hat{\bf n}{e^{ikr}\over kr}\right.\nonumber\\
&\left. +[3\hat{\bf n}(\hat{\bf n}\cdot\hat{\bf d})-\hat{\bf d}]
\bigl[ {1\over (kr)^3} - {i\over (kr)^2}\bigr]e^{ikr}
\right\}-{\hat{\bf d}\,\delta({\bf r})\over3}\,,
\label{eq:DOL}
\end{align}
where $k=\omega/c$, $\omega$ denotes the laser light frequency, and $\hat{\bf n} = {{\bf r}/ r}$.

Using the second quantized atomic field operators $\hat\psi_{g\nu}(\rv)$ and $\hat\psi_{ e\eta}(\rv)$,
the positive frequency component of
the atomic polarization density can be written in terms of contributions from different sub-level
transitions as
\begin{align}
\Pvhat^+(\rv) & = \sum_{\nu,\eta}
\Pvhat^+_{\nu\eta}(\rv)\,,
\label{pol}\\
\Pvhat^+_{\nu\eta}(\rv) &\equiv \dv_{g\nu e\eta}\hat\psi^\dagger_{g\nu}(\rv)
\hat\psi_{ e\eta}(\rv)\,,
\label{polcomp}
\end{align}
where $\dv_{g\nu e\eta}$ represents the dipole matrix element
for the transition $|e,\eta\>\rightarrow |g,\nu\>$
\beq
\dv_{g\nu e\eta}\equiv \Dc \sum_\sigma \pol_{\sigma}
\< e\eta;1g|1\sigma;g\nu\> \equiv \Dc \sum_\sigma \pol_{\sigma} {\cal C}_{\nu,\eta}^{(\sigma)}\,.
\label{dipole}
\eeq
Here the sum is over the unit circular polarization vectors $\sigma=\pm1,0$, and
${\cal C}_{\nu,\eta}^{(\sigma)}$ denote the Clebsch-Gordan coefficients of the
corresponding optical transitions.
The reduced dipole matrix element ${\cal D}$ is related to the linewidth of the transition
$\Gamma$ by
\begin{equation}
\Gamma = {{\cal D}^2\omega_0^3\over 3\pi \hbar\epsilon_0 c^3} \,,
\end{equation}
and $\dv_{e\eta g\nu}=\dv_{g\nu e\eta}^*$.
The light
fields with the polarizations $\pol_\pm$ and $\pol_0$ drive the transitions
$|g,\nu\>\rightarrow|e,\nu\pm1\>$ and $|g,\nu\> \rightarrow|e,\nu\>$,
respectively, in such a way that only the terms $\sigma=\eta-\nu$ in Eq.~\eqref{dipole}
are nonvanishing.

As described in Refs.~\cite{Ruostekoski1997,Ruostekoski2000}, in the limit of low light intensity we obtain the equation of
motion for the expectation value of the polarization
component $\Pv^+_{\nu\eta}(\rv)=\<\Pvhat^+_{\nu\eta}(\rv)\>$,
\begin{align}
  {d\over dt}\,&\Pv^+_{\nu\eta}  (\rv) =
  (i\bar{\Delta}_{g\nu e\eta}-{\Gamma\over 2})\Pv^+_{\nu\eta}(\rv)
  + i\xi \rho_\nu(\rv) {\sf P}^{\nu\eta}_{\eta\nu}
   \Dv^+_F (\rv) \nonumber\\
  & + i\xi
  \int d^3r'\,{\sf P}^{\nu\eta}_{\eta\tau}
  {\sf G}({\bf r}-{\bf r'})\, \<\hat\psi^\dagger_{g\nu}(\rv) \Pvhat^+(\rv')
  \hat\psi_{g\tau}(\rv) \> \,,
  \label{eq:CR1}
\end{align}
where we assumed that there are
no ground-state coherences between the different hyperfine states,
so that $\<\hat\psi^\dagger_{g\nu}\hat\psi_{g\tau}\> = \delta_{\nu,\tau} \rho_\nu$,
with $\rho_\nu$ the atom denisty of the ground state $|g,\nu\rangle$.

We have defined $\xi={\cal D}^2/(\hbar\eo)$ and
$\Delta_{a\nu b\eta}= \Delta_{b\eta}-\Delta_{a\nu}$ ($a,b=g,e$). In the presence of
a magnetic field $B$, the atom-light detuning is
$\Delta_{e\eta}=\omega-(\omega_0+{\mu_BB}  g_l^{(e)}\eta/\hbar)$ and
$\Delta_{g\nu}=-{\mu_BB}  g_l^{(g)}\nu/\hbar$, with $\omega$ denoting the frequency of the incident light, and $g_l^{(g)}$ and $g_l^{(e)}$ the Landé factors
of the ground and excited states, respectively.
In Eq.~\eqref{eq:CR1} we also introduced the tensor
\beq
{\sf P}^{\nu\eta}_{\mu\tau} \equiv {\dv_{g\nu e\eta}
\dv_{e\mu g\tau}\over {\cal D}^2}
= \sum_{\sigma,\varsigma}
\pol_{\sigma}\pol_{\varsigma}^* {\cal C}_{\nu,\eta}^{(\sigma)}{\cal
C}_{\tau,\mu}^{(\varsigma)}
\,.
\label{pro}
\eeq

The pair correlation function $\<\hat\psi^\dagger_{g\nu}(\rv) \Pvhat^+(\rv') \hat\psi_{g\tau}(\rv) \>$ describes light-induced correlations between
the atoms at $\rv$ and $\rv'$~\cite{Ruostekoski1997}. These correlations are non-trivial in the presence of recurrent scattering processes. In the mean-field theory
we neglect recurrent scattering by the decorrelation approximation $\<\hat\psi^\dagger_{g\nu}(\rv)\Pvhat^+(\rv')
\hat\psi_{g\tau}(\rv)\>\simeq \rho_\nu(\rv) \Pv^+(\rv')$ in Eq.~\eqref{eq:CR1}.

We are interested in the steady-state solution of  the resulting approximation to Eq.~\eqref{eq:CR1}.
In the interaction potential between the two atoms in Eq.~\eqref{eq:CR1}, we can now remove the contact term $ {\sf G}({\bf r})\rightarrow  {\sf G}({\bf r})+\delta(\rv)/3$,
and then eliminate the scattered field between Eqs.~\eqref{eq:CR1} and~\eqref{eq:MonoD} (this procedure can be derived rigorously~\cite{RuostekoskiLL1997}).
We obtain
\beq
\Pv_{\nu\eta} = \alpha_{\nu\eta}\rho_\nu {\sf P}^{\nu\eta}_{\eta\nu} (\epsilon_0 \Ev + \Pv/3)\,,
\label{simpP}
\eeq
where the polarizability is given by
\beq \label{Eq:polarizability}
\alpha_{\nu\eta}=-{\Dc^2\over \hbar \eo}{1\over \Delta_{g\nu
  e\eta}+i{\Gamma\over 2}}\,.
\eeq
Equation
\eqref{simpP} now leads to a coupled set of linear equations between the different polarization components~\cite{Ruostekoski2000} that can be solved
\beq
\Pv_{\nu\eta} = {\alpha_{\nu\eta}\rho_\nu [{\cal C}_{\nu,\eta}^{(\sigma')}]^2 \over
1- \sum_{\tau,\zeta}\alpha_{\tau\zeta}\rho_\tau [{\cal C}_{\tau,\zeta}^{(\sigma')}]^2/3} \,\epsilon_0 (\pol_{\sigma'}^*\cdot \Ev) \pol_{\sigma'} \,,
\label{soluP}
\eeq
where $\sigma'=\eta-\nu$. The sum in the denominator therefore includes the components for which ${\cal C}_{\tau,\zeta}^{(\sigma')}\neq 0$, i.e., the components for which $\Pv_{\tau\zeta}$ is parallel to $\Pv_{\nu\eta}$.

We can then separate the different vector components of the total polarization $\Pv = \sum_{\nu,\eta}\Pv_{\nu\eta}$ by considering each value of the
spherical polarization component $\sigma'$ separately
\beq
\pol_{\sigma'}^*\cdot\Pv = {\sum_{\nu,\eta} \alpha_{\nu\eta}\rho_\nu [{\cal C}_{\nu,\eta}^{(\sigma')}]^2 \over
1- \sum_{\tau,\zeta}\alpha_{\tau\zeta}\rho_\tau [{\cal C}_{\tau,\zeta}^{(\sigma')}]^2/3} \,\epsilon_0 (\pol_{\sigma'}^*\cdot \Ev)  \,.
\label{soluP2}
\eeq
This now gives the electric susceptibility for $\sigma'=\pm 1,0$.

To describe the experiment (for which the magnetic field is $B\approx 0$~G), we consider a  linearly polarized laser beam driving $\pi$ transitions ($\sigma=0$).  We also assume all the Zeeman states of the ground level $F=2$ equally populated, implying $\rho_\nu=\rho_{\rm total}/5$. In this way, the only effect of the internal structure of the atom is to multiply the polarizability given by Eq.~(\ref{Eq:polarizability}) (with $\Delta_{g\nu e\eta}=\Delta=\omega-\omega_0$) by $\sum_{\nu} [{\cal C}_{\nu,\nu}^{(0)}]^2 / 5 = 7/15$, as written in the main text.

\section{Influence of a misalignment of the probe with respect to the cloud}\label{Sec:misalignment}

\begin{figure}
\includegraphics[width=\columnwidth]{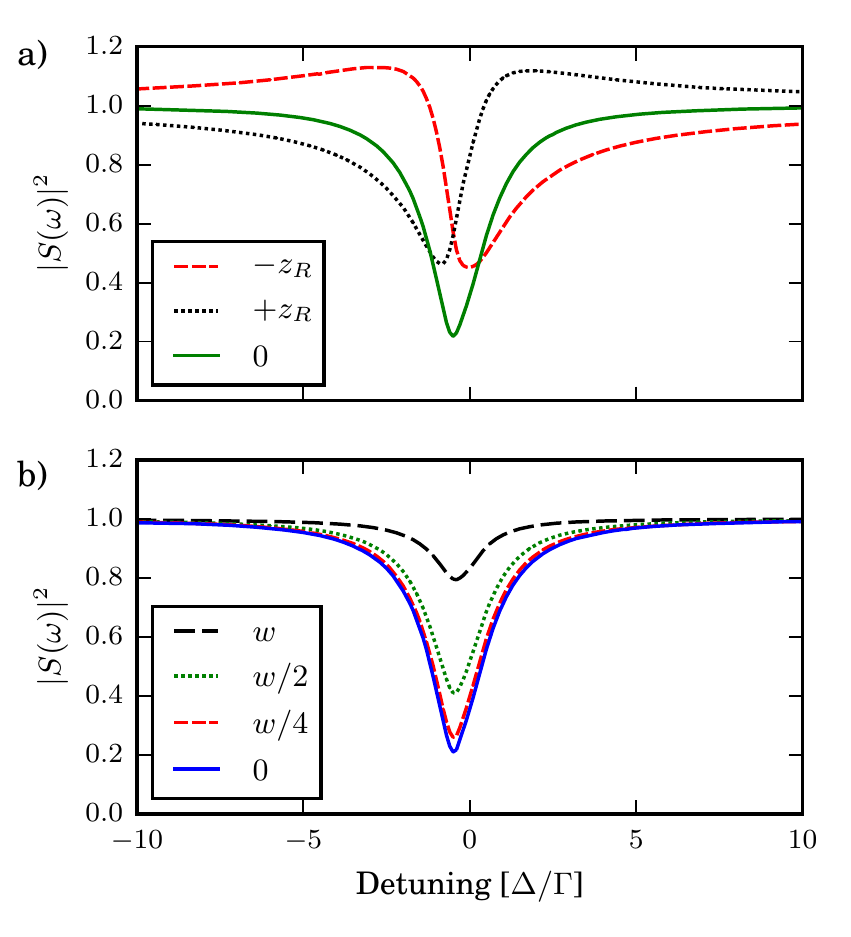}
\caption{(a) Influence of a longitudinal displacement of the position of the waist of
the probe beam with respect to the center of the atomic cloud. The focus of the beam is displaced by $z_0=\pm z_{\rm R}$, with respect to the position of the center of the cloud ($z_0=0$).
(b) Influence of a transverse misalignment of the probe beam with respect to the center of the atomic cloud. We considered 3 transverse displacements with respect to the center of the cloud ($0$): $w/4$, $w/2 $ and $w$, where $w=1.2$\,$\mu$m is the probe beam waist. Both (a) and (b) correspond to $N=20$.}
\label{fig_influence_alignment}
\end{figure}

In this last section, we analyze the influence of a misalignement of the probe with respect to the atomic cloud.
The dimensions of the cloud are $(a_\perp, a_z)=(0.2, 1.2)\,\mu$m, to be compared to the probe beam waist $w=1.20\pm0.05\,\mu$m.
These very small numbers indicate that the perfect alignment of the probe with respect to the cloud is challenging.
As demonstrated in~\cite{Roof2015}, a longitudinal displacement of the probe focal point
with respect to the center of the cloud leads to asymmetric atomic line shapes in transmission.
Reference~\cite{Roof2015} interprets this asymmetry as a lensing effect induced by the cloud.

We have modeled the effect of a misalignment of the probe using the coupled dipole equations. We calculate the transfer function
as described in the main text for different transverse and longitudinal positions of the cloud with respect to the probe beam.
Figure~\ref{fig_influence_alignment}a presents the results for a longitudinal displacement of the cloud along the direction of propagation of the probe beam for $N=20$ atoms. We observe a strong asymmetry of the line shape when the cloud is displaced by $\pm z_{\rm R}$ with respect to the position of the beam waist. This asymmetry can be understood as an effect of the  Gouy phase of the probe laser (see below).

Finally, we displace transversally the atomic cloud in the focal plane of the laser beam, perpendicular to the optical axis. The results are shown in Fig.~\ref{fig_influence_alignment}b. The main effect of the displacement is to reduce the amplitude of the line, without inducing any frequency shift or any change in the linewidth. We have
tested that this conclusion remains valid for all atom numbers investigated in this work. In particular, the transverse displacement does not induce any extra asymmetry in the line shape with respect to the case of an unshifted position of the cloud.

Both effects can be understood using the following simplified model.
Let us consider the case of a dielectric cloud with a size
smaller than $1/k$ ($k$ is the wave-vector of the light) with a polarizability $\alpha(\omega)$ (with a width $\Gamma_{\rm c}$ and a central frequency $\omega_{\rm c}$), located
at a position $(x_0,y_0,z_0)$ around the focal point of the probe beam at position $(0,0,0)$.
At a distance $z\gg |x_0|,|y_0|,|z_0|$ in the far-field,
the component of the electric field along the  $x$-axis scattered by the cloud in the direction of the propagation of the probe ($z$ axis) is~\cite{Jackson}:
\begin{eqnarray}
E_{{\rm sc},x}(z)&\approx& {k^2\over 4\pi} \alpha(\omega)|E_{\rm L}(x_0,y_0,z_0)|e^{i[\psi(z_0)+kz_0]}\\
&\times&{e^{ik\sqrt{(z-z_0)^2+x_0^2+y_0^2}}\over z}\ ,\nonumber
\end{eqnarray}
with $\psi(z_0)$ the Gouy phase given by $\psi(z_0)=-\arctan[z_0/ z_{\rm R}]$. Using the far-field expression of the laser electric field~(\ref{Eq:Gauss_beam}), we get the total field in the direction of propagation of the probe:
\begin{eqnarray}
E_{{\rm tot},x}&=&-i {z_{\rm R}\over z}|E_{\rm L }(0)|e^{ikz}\\
&\times&\left( 1+i {k^2\alpha(\omega)\over 4\pi z_{\rm R}}{|E_{\rm L}(x_0,y_0,z_0)|\over|E_{\rm L}(0)|}e^{i\psi(z_0)}\right)\ .\nonumber
\end{eqnarray}
Along the axis, the transfer function is:
\begin{equation}
{E_{{\rm tot},x}\over E_{{\rm L},x}} = 1+i {k^2\alpha(\omega)\over 4\pi z_{\rm R}}{|E_{\rm L}(x_0,y_0,z_0)|\over|E_{\rm L}(0)|}e^{i\psi(z_0)}\ ,
\end{equation}
which can be cast in the form of an on-axis transfer function:
\begin{equation}\label{Eq:S_axis(omega)}
{\cal S}_{\rm axis}(\omega) = 1-{A \over 1-2i{\omega-\omega_c \over \Gamma_c}} e^{i\psi(z_0)}\ ,
\end{equation}
with $A$ a real number
This expression shows that for a cloud located at $(x_0,y_0,0)$ (transverse displacement),
the amplitude of the extinction is attenuated
by the ratio $|E_L(x_0,y_0,z_0)|/|E_L(0)|$, as observed in Fig.~\ref{fig_influence_alignment}(b).
When the cloud is centered in  $(0,0,z_0)$ (with $z_0$ small but not negligible with respect to $z_{\rm R}$),
the phase factor $e^{i\psi(z_0)}$ involving the Gouy phase leads to an asymmetric function of $\omega$: this explains the asymmetry observed in  Fig.~\ref{fig_influence_alignment}a.
From the very slight asymmetry observed on the data (see Fig.~1a of the main text),
we infer the phase $\psi(z_0)$ by fitting the data by the expression
$|{\cal S}_{\rm axis}(\omega)|^2$ (Eq.~\ref{Eq:S_axis(omega)}). We find $\psi(z_0) \approx 0.3$\,rad and extract a longitudinal displacement of the cloud $z_0\approx 1.8\,\mu$m.
Importantly, we find that this phase is independent of the
atom number, as it should if it is a geometric phase. A lensing effect would lead to an asymmetry that would vary with the number of atoms.

We conclude from this study that a misalignment of the probe cannot introduce any narrowing of the line or any extra shift, nor wash out any double structure.
The misalignment thus cannot explain the discrepancy between the theoretical models and the data, which systematically feature a smaller shift and linewidth and no significant asymmetry.